# Tunguska: Comets, Contagion and the Vernadskiy Mission to NEA 2005NB56

Pushkar Ganesh Vaidya
Indian Astrobiology Research Centre (IARC)
pushkar@astrobiology.co.in

**ABSTRACT**

On June 30th, 1908, there was a massive explosion over Tunguska, in Central Siberia. A number of scientists have proposed that this Tunguska Phenomenon was caused due to the tangential passage of an astral body that grazed the Earths' atmosphere, underwent a partial explosion and later entered a heliocentric orbit. It has also been argued that astral bodies might deposit microbes and viruses on Earth (contributing to evolution and diseases) and may become contaminated with Earthly microbes. The identity of the Tunguska Space Body (TSB) is unknown though several likely candidates have been identified including NEA 2005NB56, a known Near Earth Asteroid (NEA). NEA 2005NB56 made a close approach to Earth when it was discovered in 2005 and will again cross Earth orbit in 2045. This gives us a unique opportunity to send a Stardustlike mission, the proposed Vernadskiy Mission, to analyze materials from NEA 2005NB56. We might be able to find some tell-tale components of Earth's atmosphere and even Earth's microorganisms incorporated in NEA 2005NB56 thereby proving beyond doubt the transfer of life from Earth to outer space.

# 1. THE TUNGUSKA SPACE BODY

On June 30th, 1908, a bollide triggered a massive explosion over Tunguska, in Central Siberia (Foschini, 2001). Over the last century several hypotheses have been put forth to explain the Tunguska Phenomenon (Foschini 2001; Napier 2009; Vasilyev 1998); one of the frontrunners includes the 1932 hypothesis of Vladimir Vernadskiy. Vernadskiy proposed that the Tunguska Phenomenon could have been due to the tangential passage of a body that grazed the Earths' atmosphere, underwent a partial explosion and later entered a heliocentric orbit (Drobyshevski et al., 2009). A part of the Tunguska Space Body (TSB) exploded at an altitude of ~5÷10 km thereby releasing energy equivalent to about 10 Mt of TNT, with the explosive power of hydrogen nuclear bombs, and leveling ~2,150 square kilometers of Siberian forest. Importantly, if the TSB explosion had occurred 4h47' later then it would have proved catastrophic for the city of St.-Petersburg and would have left over a million people dead (Lewis 1996).

It has suggested that comet Encke may have been responsible (Joseph 2009a). Comet Encke made extremely close approaches to Earth on June 16, 1908 and shed ice, rock and dust which streaked through the atmosphere of the Earth. According to Joseph, a large piece of Comet Encke may have broken off, fell to Earth, and exploded above Tunguska thereby leveling forests for hundreds of miles. Hence, comet Encke is yet another candidate for the TSB.

Two major mechanisms have been suggested to explain the breakup of the TSB. 1) It was an "explosion in flight" whereby air pressure caused the disintegration of the fast moving TSB (composed of rock and ice). 2) It was a "chemical explosion" whereby the cometary ices saturated with the solid solution of $2H_2+O_2$, the products of ice electrolysis, exploded (Drobyshevski et al., 2009).

Statistical analysis has shown a ≈83% probability that the TSB might be an asteroid and ≈17% probability of a short period (SP) comet (Farinella et al., 2001). Most researchers have been inclined to believe that the TSB was an icy body rather than a rocky asteroid which exploded in flight (Napier 2009). This is because ice breaks up easier then rocks and leaves no rocky fragments, as observed on the site of the Tunguska Phenomenon (Drobyshevski et al., 2009).

**1.1. TSB and New Explosive Cosmogony (NEC) of Minor Bodies**

The New Explosive Cosmogony (NEC) of minor bodies concept explains formation of SP cometary nuclei as due to extremely rare global explosions of thick (~800 km) electrolyzed icy envelopes of distant moonlike bodies, similar to Ganymede or Titan. The NEC holds that the Tunguska meteoroid belonged to these the SP cometary nuclei.

In the light of NEC, TSB has been deduced to be an icy cometary nucleus (whose ices contained products of its electrolysis, $2H_2+O_2$) with a mass ~ $10^{13} \div 10^{14}$ g, ~ 200÷500 m in diameter, and which was moving at a horizontal ($\delta = 0º$) velocity ~ 20÷30 km/s. NEC also helps to explain the peculiar final turn in trajectory of the TSB (by $\Delta\varphi$ ~ 10º to the west in the horizontal plane high in the atmosphere: H ≥ 10 km, possibly, even H > 50 km (Drobyshevski 2009 and refs therein).

The NEC based approach offers a satisfactory explanation of the Tunguska Phenomenon. However some minor aspects remain to be understood, like the mechanisms which produced nanodiamonds (Drobyshevski 2009).

**1.2. Return of the Tunguska Space Body**

The Tunguska Space Body (TSB) lost up to $10^{12}$ g mass during the explosive interaction with the Earths' atmosphere and managed to escape into space, falling in an orbit around the Sun. So, if the TSB has not entered the gravitational sphere of other planets then it may well return for another close encounter with Earth. Therefore it is reasonable to identify a possible candidate for the TSB (Drobyshevski et al., 2009; Foschini 2001; Lewis, 1996).

In a bold and speculative effort, the search for the TSB began amongst the Potentially Hazardous Asteroids (PHAs) for bodies which might approach Earth to within $\Delta \leq 0.05$ AU in the time period from the present to the year 2178. This was based on back integration of orbits; a TSB candidate should cross the Earth's orbit sometime around June 30, plus/minus a few days to account for the perturbations caused by other planets and probably non-gravitational forces.

Later the search for the TSB was extended to Near Earth Asteroids (NEAs). Integration of the equations of motion (based on 19-order method of Everhart) of the 6077 known NEAs back to January 1, 1908 was undertaken. The model included the effects of all large planets, Pluto and the Moon on the trajectory and orbit of this object. The coordinates of the perturbing bodies were calculated on the basis of the DE405 ephemerises (Drobyshevski et al., 2009).

Of the NEAs, 2005MB ($\Delta = 0.0971$ AU, 26.06.1908) and 2005NB56 ($\Delta = 0.06945$ AU, 27.06.1908) while of the PHAs, 25143 Itokawa ($\Delta = 0.2728$ AU, 30.06.1908) and 65909 1998 FH12 ($\Delta = 0.1746$ AU, 30.06.1908) were found to be candidates for the TSB.

Based on the data, the closest match for the TSB was NEA 2005NB56. This is because of 1) the closeness of the date (June 27, 1908) to the time of the Tunguska Phenomenon 2) closest of the four potential TSB candidates $\Delta = 0.06945$ AU 3) orbital parameters being closest to those of comet Encke and of the β-Taurid stream 4) of the four potential TSB candidates its size ($\approx 170$ m) closely matches the lower estimate derived from the angle of turn of the TSB (Drobyshevski et al., 2009 and refs therein).

NEA 2005NB56 might be the TSB. However it is not definite; for instance, it does not terminate at the exact location of the TP event unless one takes into consideration the possible effects of nongravitational forces. Interestingly, NEA 2005NB56 is slated for additional rendezvous with Earth, the first of them on July 11, 2045, at a distance of 0.04249 AU (Drobyshevski et al., 2009).

## 2. TRANSFER OF MICROORGANISMS FROM EARTH TO OUTER SPACE

The Tunguska Space Body (TSB) is the only known body to have had an extremely close interaction with Earth; just about ~5÷10 km above the Earth's surface, only to escape into space (Napier 2009). We are aware of the impact TSB left on Earth. However we are not aware of the impact the interactions left on the TSB. It would be interesting to investigate if any of Earth's atmospheric gasses or microorganisms became incorporated into the TSB when it bounced off the upper atmosphere.

It is known that microorganisms exist in significant concentrations in Earths' atmosphere (Burrows et al., 2009; Griffin 2004; Wainwright et al., 2010). Microorganisms have been found in air samples collected at heights ranging from 41 km (Wainwright et al., 2010) to 77 km (Imshenetsky 1978). The natural mechanisms which transport microorganisms to the atmosphere are storm activity, volcanic activity, monsoons, and impact events (Joseph and Schild 2010; Wainwright et al., 2010).

In fact, a particular long term study conducted close to the Tunguska Phenomenon site, in the skies of southwestern Siberia, found significant concentrations of culturable microorganisms over an altitude range of 0.5–7 km (Andreeva et al., 2002; Borodulin et al., 2005). This height range is similar to the height range of ~5÷10 km where the TSB interacted with Earth's atmosphere. Of course, the TSB-Earth interaction was fierce and explosive. However there is a possibility of contamination, either of Earth's microorganisms into the TSB, or organisms which may have resided in the TSB being deposited on Earth, or both.

**3. DISEASES FROM SPACE?**

Hoyle, Wickramasinghe, Napier and others (Hoyle and Wickramasinghe 2000, Napier and Wickramasinghe 2010; Wickramasinghe 2010; Wickramasinghe et al., 2009) have provided considerable evidence indicating microbial life can flourish within the heart of comets and may be deposited on other planets including Earth. Wainwright et al., (2010), based on evidence of bacteria in the upper atmosphere and stratosphere, argue that life may be continually incoming from space and outgoing from Earth. Joseph (2009b; Joseph and Schild 2010) has developed a detailed model explaining how microbes can be lofted into the stratosphere and periodically ejected into space during particularly powerful solar storms. Joseph (2009b; Joseph and Schild 2010) has also provided and reviewed evidence that microbes can travel from planet to planet and solar system to solar system encased in asteroids, comets and other stellar debris, and that they can survive the impact and heat of ejection and reentry into the atmosphere. Therefore, there is good reason to suspect that a stellar object striking and skimming along the upper atmosphere could eject microbes into the atmosphere of Earth and become contaminated with microbes already present at these heights.

Joseph (2000, 2009c; Joseph and Schild 2010) has developed a detailed genetic model of cosmic evolution, and has argued that as microbes and viruses are transferred from planet to planet, they exchange and acquire DNA. According to Joseph (2000, 2009c) these genes were then transferred to the eukaryotic genome, contributing to the evolution of multi-cellular life leading to humans. Likewise, Wainwright et al., (2010) propose that incoming and outgoing microbes may exchange DNA via horizontal gene transfer, and this genetic exchange contributes to the evolution of life on Earth.

Wickramasinghe (2010) also believes that space-traveling microbes and viruses contribute to evolution, but are also responsible for "errors" introduced into the genome. These "errors" result in disease, disability, and death. Hoyle and Wickramasinghe (1979, 1986) have referred to this as "diseases from space."

Is there any evidence that microbes and viruses were deposited on Earth by the TSB or following the Tunguska impact in 1908? The evidence is indirect at best, i.e. the 1918 flu epidemic which killed over 20 million people world wide (Joseph 2009a). It is unknown if the TSB is comet Encke. However, Comet Encke made extremely close approaches to Earth on June 16, 1908, and again on October 27 1914, and was at perihelion on 1918. And with each approach, Comet Encke shed ice, rock and dust which streaked through the atmosphere of Earth.

According to Hoyle and Wickramasinghe (1979, 1986, 2000; Wickramasinghe 2010; Wickramasinghe et al., 2009) under these circumstances microbes and viruses would be shed from the comet and would be deposited on Earth and could induce disease. Wainwright et al., (2010) argue that microbes from space would exchange DNA with microbes of Earth, and the same has been said of viruses (Joseph and Schild 2010). There is also considerable evidence that the 1918 flu epidemic was due to gene mixing between an already established virus and a completely unknown "new" virus (Joseph 2009a).

As detailed by Joseph (2009a):

> "In 2005, scientists from the Armed Forces Institute of Pathology in Washington, D.C., resurrected the 1918 virus from bodies that had been preserved in the permanently frozen soil of Alaska. They soon discovered that a completely new virus had combined with an old virus, exchanging and recombining genes, creating a hybrid that transformed mild strains of the flu virus into forms far more deadly and pathogenic. They also confirmed that the 1918 Spanish flu virus originated in the sky, first infecting birds and then spreading and proliferating in humans."

**4. VERNADSKIY MISSION**

NEA 2005NB56 has been identified as a close match to the TSB. It will make its next close approach to Earth in 2045. This provides us a unique opportunity to send a mission to NEA 2005NB56. This proposed mission, tentatively called as Vernadskiy Mission, should have a design similar to the Stardust mission which captured material from comet 81P/Wild 2 in 2004, which was subsequently analyzed (Brownlee et al., 2006).

Analysis of material from NEA 2005NB56 will possibly help us to establish if NEA 2005NB56 is indeed the TSB and if it played any role in "diseases from space." The scientific community will have to identify and search for some tell-tale components of Earth's atmospheric components and also for microorganisms and viruses, viable or otherwise.

**5. CONCLUSIONS**

Vernadskiy Mission will help us analyze the material from NEA 2005NB56. If we are able to establish that NEA 2005NB56 is indeed the TSB then one of the greatest mysteries of our times will be solved. The bonus might be that we may find microorganisms. This would for the first time establish beyond doubt the transfer of life via mechanisms of panspermia and may shed light on the possibility that diseases may originate in space.

**ACKNOWLEDGMENTS**

The author is grateful to his friend, Kapil Mayekar, for his cushioning support.